\documentstyle[12pt]{article}

  \def\pp{{\mathchoice
          {
              \kern 1pt%
              \raise 1pt
              \vbox{\hrule width5pt height0.4pt depth0pt
                    \kern -2pt
                    \hbox{\kern 2.3pt
                          \vrule width0.4pt height6pt depth0pt
                          }
                    \kern -2pt
                    \hrule width5pt height0.4pt depth0pt}%
                    \kern 1pt
           }
            {
              \kern 1pt%
              \raise 1pt
              \vbox{\hrule width4.3pt height0.4pt depth0pt
                    \kern -1.8pt
                    \hbox{\kern 1.95pt
                          \vrule width0.4pt height5.4pt depth0pt
                          }
                    \kern -1.8pt
                    \hrule width4.3pt height0.4pt depth0pt}%
                    \kern 1pt
            }
            {
              \kern 0.5pt%
              \raise 1pt
              \vbox{\hrule width4.0pt height0.3pt depth0pt
                    \kern -1.9pt  
                    \hbox{\kern 1.85pt
                          \vrule width0.3pt height5.7pt depth0pt
                          }
                    \kern -1.9pt
                    \hrule width4.0pt height0.3pt depth0pt}%
                    \kern 0.5pt
            }
            {
              \kern 0.5pt%
              \raise 1pt
              \vbox{\hrule width3.6pt height0.3pt depth0pt
                    \kern -1.5pt
                    \hbox{\kern 1.65pt
                          \vrule width0.3pt height4.5pt depth0pt
                          }
                    \kern -1.5pt
                    \hrule width3.6pt height0.3pt depth0pt}%
                    \kern 0.5pt
            }
        }}

  \def\mm{{\mathchoice
                       {
                             \kern 1pt
               \raise 1pt    \vbox{\hrule width5pt height0.4pt depth0pt
                                  \kern 2pt
                                  \hrule width5pt height0.4pt depth0pt}
                             \kern 1pt}
                       {
                            \kern 1pt
               \raise 1pt \vbox{\hrule width4.3pt height0.4pt depth0pt
                                  \kern 1.8pt
                                  \hrule width4.3pt height0.4pt depth0pt}
                             \kern 1pt}
                       {
                            \kern 0.5pt
               \raise 1pt
                            \vbox{\hrule width4.0pt height0.3pt depth0pt
                                  \kern 1.9pt
                                  \hrule width4.0pt height0.3pt depth0pt}
                            \kern 1pt}
                       {
                           \kern 0.5pt
             \raise 1pt  \vbox{\hrule width3.6pt height0.3pt depth0pt
                                  \kern 1.5pt
                                  \hrule width3.6pt height0.3pt depth0pt}
                           \kern 0.5pt}
                       }}

\catcode`@=11
\def\un#1{\relax\ifmmode\@@underline#1\else
        $\@@underline{\hbox{#1}}$\relax\fi}
\catcode`@=12

\let\du=\du                     

\def\a{\alpha}

\def\c{\chi}
\def\d{\delta}

\def\f{\phi}

\def\h{\eta}

\def\j{\psi}

\def\m{\mu}
\def\n{\nu}

\def\p{\pi}
\def\q{\theta}
\def\r{\rho}
\def\s{\sigma}

\def\ve{\varepsilon}

\def\bo{{\raise-.5ex\hbox{\large$\Box$}}}               
\def\pa{\partial}                                       
\def\pr{\prod}                                          
\def\TH{{\raise.2ex\hbox{$\displaystyle \bigodot$}\mskip-4.7mu \llap H \;}}
\def\face{{\raise.2ex\hbox{$\displaystyle \bigodot$}\mskip-2.2mu \llap {$\ddot
        \smile$}}}                                      

\def\sp#1{{}^{#1}}                              
\def\abs#1{\left| #1\right|}                    
\def\leftrightarrowfill{$\mathsurround=0pt \mathord\leftarrow \mkern-6mu
        \cleaders\hbox{$\mkern-2mu \mathord- \mkern-2mu$}\hfill
        \mkern-6mu \mathord\rightarrow$}
\def\dvec#1{\vbox{\ialign{##\crcr
        \leftrightarrowfill\crcr\noalign{\kern-1pt\nointerlineskip}
        $\hfil\displaystyle{#1}\hfil$\crcr}}}           

\def\frac#1#2{{\textstyle{#1\over\vphantom2\smash{\raise.20ex
        \hbox{$\scriptstyle{#2}$}}}}}                   
\def\sfrac#1#2{{\vphantom1\smash{\lower.5ex\hbox{\small$#1$}}\over
        \vphantom1\smash{\raise.4ex\hbox{\small$#2$}}}} 
\def\bfrac#1#2{{\vphantom1\smash{\lower.5ex\hbox{$#1$}}\over
        \vphantom1\smash{\raise.3ex\hbox{$#2$}}}}       
\def\afrac#1#2{{\vphantom1\smash{\lower.5ex\hbox{$#1$}}\over#2}}    

\def\[{\lfloor{\hskip 0.35pt}\!\!\!\lceil}
\def\]{\rfloor{\hskip 0.35pt}\!\!\!\rceil}
\def\Lag{{\cal L}}
\def\du#1#2{_{#1}{}^{#2}}

\def\fracm#1#2{\hbox{\large{${\frac{{#1}}{{#2}}}$}}}

\def\un{\underline}
\def\fracmm#1#2{{{#1}\over{#2}}}

\def\low#1{{\raise -3pt\hbox{${\hskip 0.75pt}\!_{#1}$}}}

\newskip\humongous \humongous=0pt plus 1000pt minus 1000pt
\def\caja{\mathsurround=0pt}
\def\eqalign#1{\,\vcenter{\openup2\jot \caja
        \ialign{\strut \hfil$\displaystyle{##}$&$
        \displaystyle{{}##}$\hfil\crcr#1\crcr}}\,}
\newif\ifdtup

\def\ref#1{$\sp{#1)}$}

\def\pl#1#2#3{Phys.~Lett.~{\bf {#1}B} (19{#2}) #3}
\def\np#1#2#3{Nucl.~Phys.~{\bf B{#1}} (19{#2}) #3}

\def\pr#1#2#3{Phys.~Rev.~{\bf D{#1}} (19{#2}) #3}
\def\cqg#1#2#3{Class.~and Quantum Grav.~{\bf {#1}} (19{#2}) #3}

\parskip=\medskipamount                 
\thicklines                         

\begin{document}


\thispagestyle{empty}               

\def\border{                                            
        \setlength{\unitlength}{1mm}
        \newcount\xco
        \newcount\yco
        \xco=-24
        \yco=12

        \par\vskip-8mm}

\def\headpic{                                           
        \indent
        \setlength{\unitlength}{.8mm}
        \thinlines
        \par
        \par\vskip-6.5mm
        \thicklines}

\border\headpic {\hbox to\hsize{
\vbox{\noindent  ITP--UH--02/00 \hfill February 2000 \\
NBI--HE--00--05 \hfill hep-th/0002134 \\ }}}

\noindent
\vskip1.3cm
\begin{center}

{\Large\bf  Manifestly N=3 Supersymmetric Euler-\vglue.1in
          Heisenberg Action in Light-Cone Superspace}

\vglue.3in

Thomas B\"ottner, Sergei V. Ketov\footnote{Also at NBI, Univ. of Copenhagen, 
Denmark, and HCEI, Academy of Sciences, Tomsk, Russia} and Thomas Lau

{\it Institut f\"ur Theoretische Physik, Universit\"at Hannover}\\
{\it Appelstra\ss{}e 2, Hannover 30167, Germany}\\
{\sl boettner,ketov,lau@itp.uni-hannover.de}

\end{center}

\vglue.3in

\begin{center}
{\Large\bf Abstract}
\end{center}

\noindent

We find a manifestly N=3 supersymmetric generalization of the four-dimensional
Euler-Heisenberg (four-derivative, or $F^4$) part of the Born-Infeld action in 
light-cone gauge, by using N=3 light-cone superspace. 

\newpage

\section{Introduction}

The {\it Born-Infeld} (BI) action in flat spacetime,~\footnote{We use 
$\h_{\m\n}={\rm diag}(+,-,-,-)$ and $\hbar=c=1$.}
$$ S_{\rm BI}= \fracmm{1}{b^2}\int d^4x \left\{ 1-\sqrt{-\det(\h_{\m\n}+
bF_{\m\n})}\right\}~, \eqno(1.1) $$
is the particular non-linear generalization of Maxwell theory, $F_{\m\n}=
\pa_{\m}A_{\n}-\pa_{\n}A_{\m}$. The action (1.1) was initially introduced 
to regularize both the electric field and the self-energy of a point-like 
charge in electrodynamics \cite{bi}. Much later, the BI action was recognized 
as the leading contribution to the effective action of open strings in an
abelian background with constant field strength $F$ \cite{bis}, and as the
essential part of the D3-brane action as well \cite{bra}, with $b=2\p\a'$. 
The action (1.1) has many remarkable properties, e.g., causal propagation 
and electric-magnetic duality \cite{cas, dual}. 

The BI Lagrangian can be rewritten to the form
$$ L=-\fracm{1}{2}p^{\m\n}F_{\m\n} + H(P,Q)~,\eqno(1.2)$$
where the auxiliary antisymmetric tensor $p_{\m\n}$ and  the BI structure 
function 
$$ H(P,Q)= \fracmm{1}{b^2}\left( 1-\sqrt{1-2b^2P+b^4Q^2}\right)~,\eqno(1.3)$$ 
as well as the definitions 
$$P=\fracmm{1}{4}p_{\m\n}p^{\m\n}~,\quad 
Q=\fracmm{i}{4}p_{\m\n}\tilde{p}^{\m\n}~,\qquad
 \tilde{p}^{\m\n}=\fracmm{1}{2}\ve^{\m\n\r\s}p_{\r\s}~,\eqno(1.4)$$
have been introduced. Eliminating $p_{\m\n}$ from eq.~(1.2) results in the 
 equivalent Lagrangian 
$$ L =\fracmm{1}{b^2}\left[ 1-\sqrt{1+\fracm{b^2}{2}F^2-\fracm{b^4}{16}
(F\tilde{F})^2}\,\right]~,\eqno(1.5)$$
where we have defined $F^2=F^{\m\n}F_{\m\n}\,$, 
$\tilde{F}^{\m\n}=\fracm{1}{2}\ve^{\m\n\r\s}F_{\r\s}$ ~and~ 
$F\tilde{F}=F^{\m\n}\tilde{F}_{\m\n}\,$.

Supersymmetric generalizations of the BI action are of particular interest in
connection to superstring theory (see ref.~\cite{tser} for a recent review). 
The super-BI actions describing D-branes can be naturally interpreted as the 
Goldstone-type actions associated with partial supersymmetry breaking, while 
they can still be duality invariant too. The manifestly N=1 supersymmetric 
generalization of the four-dimensional BI action in N=1 superspace was 
discovered long time ago \cite{cf} (see also ref.~\cite{pu}), while its 
manifestly N=2 supersymmetric generalization in N=2 superspace was found only
recently \cite{ket} (see ref.~\cite{kt} too). To our knowledge, the higher 
(N=3 or N=4) manifestly supersymmetric generalizations of the four-dimensional
bosonic BI action (1.1) are not known in any form.

Supersymmetry apparently prefers the parametrization of the BI action in
terms of the Maxwell term $L_2=-\fracm{1}{4}F^2$ and the Maxwell stress-energy
tensor squared \cite{ket}, 
$$L_4=\fracm{1}{32}\left\{ (F^2)^2+(F\tilde{F})^2\right\}=\fracm{1}{8}
(F^+)^2(F^-)^2~,\quad F^{\pm}_{\m\n}=\fracm{1}{2}\left(F_{\m\n}\pm i
\tilde{F}_{\m\n}\right)~.\eqno(1.6)$$
This term is known as the {\it Euler-Heisenberg} (EH) Lagrangian \cite{eu}. 
The EH action also appears as the bosonic part of the one-loop effective 
action in N=1 supersymmetric scalar electrodynamics with the parameter 
$b^{-1}=2\sqrt{6}\p m^2/e^2$. One easily finds that
$$ L_{\rm BI}=\fracmm{1}{b^2}\left\{ 1-\sqrt{(1-b^2L_2)^2-2b^4L_4}\,\right\}=
L_2+b^2L_4+O(F^6)~.\eqno(1.7)$$
A manifestly N=4 supersymmetric generalization  of the BI action is known to
be the formidable problem, though it is highly desirable, e.g., for an 
investigation of quantum properties of D3-branes and their comparison to 
supergravity \cite{tro,mil}. Even the $N>2$ supersymmetrization of the EH-term
$L_4$, representing the four-derivative terms $(F^4)$, is non-trivial. The 
additional terms with four derivatives in the N=4 BI action were determined in
 ref.~\cite{liu} in N=1 superspace, by imposing the $SU(3)$ internal symmetry 
on three N=1 chiral multiplets extending an N=1 (abelian) vector multiplet to 
an N=4 vector multiplet, with manifest (linearly realised) N=1 off-shell 
supersymmetry. The manifestly N=2 supersymmetric form of the N=4 EH action was
derived in ref.~\cite{stony} in N=2 projective superspace, while its equations
of motion can also be written in terms of on-shell N=4 superfields in
harmonic superspace \cite{hw}. It is the purpose of this Letter to write
down an off-shell, manifestly N=3 supersymmetric formulation of the N=4 EH
action in N=3 light-cone superspace. 

Our paper is organized as follows. In sect.~2 we introduce a light-cone
gauge and rewrite the EH Lagrangian in terms of physical (transverse) degrees 
of freedom up to the relevant order. In sect.~3 we introduce N=3 light-cone
superspace and deduce an N=3 supersymmetric generalization of the EH action
in terms of a single N=3 light-cone superfield. The obstructions encountered 
in our efforts to find a similar, manifestly N=4 supersymmetric EH action 
in N=4 light-cone superspace are discussed in Conclusion (sect.~4).  

\section{EH action in light-cone gauge}

The light-cone formulation of a gauge theory (in light-cone gauge) keeps only
physical (transverse) degrees of freedom in the field theory by giving up its
manifest Lorentz invariance. The light-cone formulation is, therefore, very 
suitable for an off-shell formulation of N-extended supersymmetric gauge field 
theories with manifest supersymmetry in N-extended light-cone superspace 
\cite{bri,man,mk}. 

We define light-cone coordinates in Minkowski spacetime as 
$$\eqalign{
x^+=\fracmm{1}{\sqrt{2}}\left( x^0+ x^3\right)~,\quad & \quad
x^-=\fracmm{1}{\sqrt{2}}\left( x^0- x^3\right)~,\cr
x=\fracmm{1}{\sqrt{2}}\left( x^1+ ix^2\right)~,\quad & \quad 
\bar{x}~ =\fracmm{1}{\sqrt{2}}\left( x^1- ix^2\right)~,\cr}\eqno(2.1)$$
and similarly for the gauge vector field, $A_{\m}\to (A^+,A^-,A,\bar{A})$.
The real coordinate $x^+$ is going to be considered as `light-cone time'. 
The linear transformation (2.1) of spacetime coordinates is obviously 
non-singular (with the Jacobian equal to $i$), while it does not preserve 
the Minkowski metric (i.e. it is not a Lorentz-transformation). 
   
The light-cone gauge reads
$$ A^+=0~.\eqno(2.2)$$
In this (physical) gauge the $A^-$ component of the gauge field $A_{\m}$ is 
supposed to be eliminated via its (non-dynamical) equation of motion, whereas
the transverse components $(A,\bar{A})$ are supposed to represent the physical
propagating fields. 

It is easy to solve the equation of motion for $A^-$ in the Maxwell theory, 
where it takes the form of a linear equation in the light-cone gauge 
({\it cf.} refs.~\cite{bri,man,mk}). It becomes, however, a highly non-trivial 
problem in the BI or EH theory, where it takes the form of a non-linear partial
differential equation. The equations of motion amount to the conservation 
law for the $p$-tensor,
$$ \pa^{\m}p_{\m\n}=0~,\eqno(2.3)$$
while the $p_{\m\n}$ in the BI theory is given by 
$$ p_{\m\n}=\fracmm{ b^2F_{\m\n} -\fracm{b^4}{4}(F\tilde{F})\tilde{F}_{\m\n}}{
\sqrt{ 1+ \fracm{b^2}{2}F^2-\fracm{b^4}{16}(F\tilde{F})^2}}~.\eqno(2.4)$$
By the use of the Bianchi identity, $\pa^{\m}\tilde{F}_{\m\n}=0$, we find the
following equation for $A^-$:
$$\eqalign{
\pa^{\m}F_{\m-}~=~&b^2\left\{ -\frac{1}{2}\pa^{\m}F_{\m-}F^2+\frac{1}{4}
\tilde{F}_{\m-}\pa^{\m}(F\tilde{F})+\frac{1}{4}F_{\m-}\pa^{\m}F^2\right\}\cr
& +b^4\left\{ -\frac{1}{16}(F\tilde{F})\tilde{F}_{\m-}\pa^{\m}F^2+
\frac{1}{16}\pa^{\m}F_{\m-}(F\tilde{F})^2 \right. \cr
& \left. +\frac{1}{16}\tilde{F}_{\m-}\pa^{\m}(F\tilde{F})F^2-\frac{1}{32}
F_{\m-}\pa^{\m}(F\tilde{F})^2\right\}~. \cr} \eqno(2.5)$$
We use a perturbative {\it Ansatz}, in powers of the small parameter 
$b^2$, for a solution to eq.~(2.5),
$$ A^-(x)=\sum^{\infty}_{n=0}b^{2n}A^-_{(2n)}(x)~.\eqno(2.6)$$
As regards the leading and sub-leading terms, we find
$$   \eqalign{
A^-_{(0)}~=~ & \fracmm{1}{\pa^+}\left(\bar{\pa}A+\pa\bar{A}\right)~,\cr
A^-_{(2)}~=~ &\left. \fracmm{1}{(\pa^+)^2}\left[ -\frac{1}{2}\pa^{\m}F_{\m-}F^2
+\frac{1}{4}\tilde{F}_{\m-}\pa^{\m}(F\tilde{F})+\frac{1}{4}F_{\m-}\pa^{\m}F^2
\right] \right|_{A^-=A^-_{(0)}}~,\cr}\eqno(2.7)$$
where we have used the notation $\pa^+=\pa/\pa x^-$. The multiple factors 
$(\pa^+)^{-1}$ in our actions are harmless after rewriting them to momentum
space. The first line of eq.~(2.7) coincides with the exact solution in the 
Maxwell theory. 

According to eq.~(1.7), the EH term $L_4$ is the leading $b^2$-correction to 
the Maxwell term $L_2$ in the BI theory.  The light-cone formulation of the BI
Lagrangian in the same approximation is thus given by the terms written down 
on the right-hand-side of eq.~(1.7) after a substitution of eq.~(2.2) and 
the first line of eq.~(2.8). After some algebra and partial integration we 
find
$$ \eqalign{
L[A,\bar{A}] & ~= -\fracmm{1}{4}F^2 + \fracmm{b^2}{8}(F^+)^2(F^-)^2 \cr
& ~= -A\bo\bar{A}+ 2b^2\abs{(\pa\bar{A})^2+\pa^+\bar{A}\fracmm{\bo}{2\pa^+}A
-\pa^+\bar{A}\fracmm{\pa^2}{\pa^+}\bar{A}}^2 +O(b^4)~,\cr}\eqno(2.8)$$
where we have used the notation $\pa=\pa/\pa x$ and  $\bar{\pa}=
\pa/\pa\bar{x}$. Eq.~(2.8) can be thought of as the light-cone EH Lagrangian.
Its N=3 supersymmetrization is discussed in the next sect.~3.

\section{N=3 light-cone superspace action}

The light-cone N=3 supersymmetry algebra reads
$$  \{Q^m,\bar{Q}_n\}=-\sqrt{2}\d^m_nP^+~,\qquad m,n=1,2,3~,\eqno(3.1)$$
where the supersymmetry charges $Q^r$ transform in the fundamental 
representation of $SU(3)$. A natural representation of the algebra (3.1) in 
N=3 light-cone superspace $Z=(x^{\m},\q^m,\bar{\q}_n)$ is given by
$$\eqalign{  
Q^m~=~& -\fracmm{\pa}{\pa\bar{\q}_m}+\fracmm{i}{\sqrt{2}}\q^m\pa^+~,\cr
\bar{Q}_n~=~&\fracmm{\pa}{\pa\q^n}-\fracmm{i}{\sqrt{2}}\bar{\q}_n\pa^+~.\cr}
\eqno(3.2)$$
The covariant derivatives in N=3 light-cone superspace are
$$\eqalign{
D^m~=~&-\fracmm{\pa}{\pa\bar{\q}_m}-\fracmm{i}{\sqrt{2}}\q^m\pa^+~,\cr
\bar{D}_n~=~&\fracmm{\pa}{\pa\q^n}+\fracmm{i}{\sqrt{2}}\bar{\q}_n\pa^+~.\cr}
\eqno(3.3)$$
They anticommute with the supersymmetry charges (3.2) and obey the same 
algebra (3.1). The irreducible off-shell representations of N=3 light-cone 
supersymmetry are easily obtained by imposing the covariant chirality 
condition on N=3 light-cone superfields $\f(Z)$,
$$ D^m\f(Z)=0~.\eqno(3.4)$$
A solution to eq.~(3.4) in components is just given by an arbitrary complex
function $\f(x^+,x^- +\fracmm{i}{\sqrt{2}}\q^m\bar{\q}_m,x,\bar{x};\q^n)
\equiv \f(y;\q)$. Its expansion in the chiral superspace reads
$$ \f(y;\q)=
\fracmm{1}{\pa^+}A(y)+\fracmm{i}{\pa^+}\q^m\bar{\c}_m(y) +\fracmm{i}{2}
\q^m\q^n\ve_{mnp}C^p(y) +\fracmm{1}{3!}\ve_{mnp}\q^m\q^n\q^p\j(y)~.\eqno(3.5)$$
The light-cone N=3 supersymmetry transformation laws for the components are
$$ \eqalign{
\d A~=~ & i\ve^n\bar{\c}_n~,\cr
\d\bar{\c}_m~=~&\sqrt{2}\bar{\ve}_m\pa^+A +\ve_{mnp}\ve^n\pa^+C^p~,\cr
\d C^p~=~& -i\sqrt{2}\ve^{pqr}\bar{\ve}_q\bar{\c}_r-i\ve^p\j~,\cr
\d\j~=~& -\sqrt{2}\bar{\ve}_n\pa^+C^n~,\cr}\eqno(3.6)$$
where $(\ve^n,\bar{\ve}_m)$ are the infinitesimal anticommuting parameters.

All our field components have canonical dimensions. The complex field $A$ can
be identified with the physical (translational) vector field components, the 
spinors $\bar{\c}_m$ in the fundamental representation ${\bf 3}$ of $SU(3)$ 
with a triplet of photinos, the singlet spinor $\j$ with extra photino, and 
the complex triplet $C^m$ with Higgs fields in ${\bf 3}$ of $SU(3)$. The 
physical content thus coincides with that of the N=4 supersymmetric abelian 
vector multiplet having a single photon field, photinos in the fundamental 
representation ${\bf 4}$ of $SU(4)$ and Higgs fields in real ${\bf 6}$ of 
$SU(4)$, after their decomposition with respect to the $SU(3)$ subgroup of the
internal symmetry $SU(4)$. This is the manifestation of the well-known fact 
that N=3 and N=4 supersymmetric vector multiplets are physically equivalent.

It is now straightforward (though very tedious) to find the N=3 supersymmetric
generalization of the bosonic EH light-cone action (2.8) in N=3 light-cone
superspace,
$$ S = \int d^4x d^3\q d^3\bar{\q}\,\Lag(\f,\bar{\f})
=- \int d^4x (D)^3 (\bar{D})^3\,\Lag(\f,\bar{\f})
~,\eqno(3.7)$$
where $(D)^3=\ve_{mnp}D^mD^nD^p$ and similarly for $(\bar{D})^3$.  After some 
trials and errors, we find
$$\eqalign{
36(-i\sqrt{2})^3\Lag(\f,\bar{\f})= & -\f\fracmm{\bo}{\pa^+}\bar{\f}
+2b^2\left\{ \fracmm{1}{\pa^{+3}}\left(\bar{\pa}\pa^+\f\bar{\pa}\pa^+\f\right)
(\pa\pa^+\bar{\f})^2
\right. \cr
& +\fracmm{1}{\pa^{+3}}\left( \pa^{+2}\f\bar{\pa}^2\f\right)
\pa^{+2}\bar{\f}\pa^2\bar{\f} +
\fracmm{1}{2\pa^+}(\f) (\pa\pa^+\bar{\f})^2\bo\bar{\f}
\cr
& +\fracmm{1}{4\pa^{+3}}\left(\pa^{+2}\f\bo\f\right)\pa^{+2}\bar{\f}\bo\bar{\f}
- \fracmm{1}{\pa^{+3}}\left(\pa^{+2}\f\bar{\pa}^2\f\right)
(\pa\pa^{+}\bar{\f})^2 \cr
& -\fracmm{1}{2\pa^{+3}}\left(\pa^{+2}\f\bo\f\bar{\pa}^2\f\right)
\pa^{+2}\bar{\f} -
 \fracmm{1}{\pa^{+3}}\left(\bar{\pa}\pa^{+}\f\bar{\pa}\pa^{+}\f\right)
\pa^{+2}\bar{\f}\pa^2\bar{\f} \cr
 &\left. +\fracmm{1}{2\pa^{+3}}\left(\bar{\pa}\pa^{+}\f\bar{\pa}\pa^{+}\f
\bo\f\right)\pa^{+2}\bar{\f}
- \fracmm{1}{2\pa^+}(\f)\pa^{+2}\bar{\f}\bo\bar{\f}
\pa^2\bar{\f}\right\}~.\cr}\eqno(3.8)$$
The bosonic part of this action is given by
$$ \eqalign{
\Lag_{\rm bos.}= & 
 -A\bo\bar{A}+ 2b^2\abs{(\pa\bar{A})^2+\pa^+\bar{A}\fracmm{\bo}{2\pa^+}A
-\pa^+\bar{A}\fracmm{\pa^2}{\pa^+}\bar{A}}^2 \cr
& -\fracmm{1}{2}C^p\bo\bar{C}_p -2b^2\left\{ \fracmm{2}{\pa^{+2}} \left(
\bar{\pa}\pa^+C^m\bar{\pa}\pa^+A\right)\left(\pa\pa^+\bar{C}_m\pa\pa^+\bar{A}
\right)\right. \cr
& +\fracmm{1}{2\pa^{+2}}\left(\pa^{+2}C^p\bar{\pa}^2A+\bar{\pa}^2C^p\pa^{+2}A
\right)\left(\pa^{+2}\bar{C}_p\pa^2\bar{A}+\pa^2\bar{C}_p\pa^{+2}\bar{A}\right)
\cr
& +\fracmm{1}{8\pa^{+2}}\left(\pa^{+2}C^p\bo A+\bo C^p\pa^{+2}A\right)
\left(\pa^{+2}\bar{C}_p\bo\bar{A}+\bo\bar{C}_p\pa^{+2}\bar{A}\right)\cr
& +\left[ \fracmm{1}{4}C^p\left(2\pa\pa^+\bar{C}_p\pa\pa^+\bar{A}\bo\bar{A}
+\bo\bar{C}_p\pa\pa^+\bar{A}\pa\pa^+\bar{A}\right) \right.\hspace{4.3cm}(3.9)
 \cr
& -\fracmm{1}{\pa^{+2}}\left(\pa^{+2}C^p\bar{\pa}^2A+\bar{\pa}^2C^p\pa^{+2}A
\right)\pa\pa^+\bar{C}_p\pa\pa^+\bar{A}\cr
& \left.\left. -\fracmm{1}{4\pa^{+2}}\left(\pa^{+2}C^p\bo A\bar{\pa}^2A
+\bar{\pa}^2C^p\pa^{+2}A\bo A +\bo C^p\bar{\pa}^2A\pa^{+2}A\right)\pa^{+2}
\bar{C}_p +{\rm h.c.} \right]\right\} ~.\cr}$$

One of the obvious features of both eqs.~(3.8) and (3.9) is the apparent 
presence of higher derivatives, as may have been expected from the experience
with the manifestly N=2 supersymmetric generalization of the BI action in the
covariant N=2 superspace \cite{ket}. The expected correspondence to the 
component D3-brane effective action having non-linearly realized extended
supersymmetry and no higher derivatives implies the existence of a field
redefinition that would eliminate the higher-derivative terms in our action
and make its N=3 supersymmetry to be non-linearly realised (i.e non-manifest)
\cite{tser}.  We also note the absence of quartic $(C^4)$ scalar terms 
and the on-shell ($\bo A=\bo C=0$) invariance of our action under constant 
shifts, $C_p(x)\to C_p(x)+c_p$, which are supposed to be related to the 
possible interpretation of the $C_p$ fields as the Goldstone scalars 
associated with spontaneoulsy broken translations in the full N=3 BI action.

\section{Conclusion}

Our main results are given by eqs.~(2.8), (3.8) and (3.9). Our initial 
motivation was to construct an N=4 supersymmetric generalization of the EH
action in the light-cone gauge. The N=4 light-cone supersymmetry algebra is
given by eq.~(3.1), where the indices now take four values. Equations (3.2),  
(3.3) and (3.4) are still valid in N=4 light-cone superspace, where they have 
to supplemented by an extra (generalized reality) condition \cite{bri},
$$ D^mD^n\bar{\f}=\fracmm{1}{2}\ve^{mnpq}\bar{D}^p\bar{D}_q\f,~~{\rm or,~
equivalently,}~~ \bar{\f}=\fracmm{1}{48\pa^{+2}}\ve^{mnpq}\bar{D}_m\bar{D}_n
\bar{D}_p\bar{D}_q\f~.\eqno(4.1)$$
The restricted chiral N=4 light-cone superfield $\f$ is equivalent to the
chiral N=3 superfield in eq.~(3.5). Our efforts to construct an N=4 
generalization of eq.~(2.8)  along the similar lines (sect.~3) unexpectedly
failed, while eq.~(4.1) was the main obstruction. We conclude that even a 
manifestly N=4 supersymmetric generalization of the EH action in the 
light-cone gauge seems to be highly non-trivial, if any, not to mention an 
even more ambitious (manifest) N=4 supersymmetrization of the BI action.

\section*{Acknowledgements}

We are grateful to Norbert Dragon, Gordon Chalmers, Olaf Lechtenfeld and 
Daniela Zanon for useful discussions.

\end{document}